\documentclass[amsmath, amssymb, twocolumn, aps, prx, superscriptaddress, footinbib]{revtex4-1}

\usepackage{mathtools}
\usepackage{mathrsfs, amsmath, wasysym}
\usepackage{bbold}
\usepackage[mathscr]{eucal}
\usepackage{xfrac}
\usepackage{graphicx}
\usepackage{dcolumn}
\usepackage{bm}
\usepackage{color}
\usepackage{tabularx}
\usepackage{natbib}
\usepackage[colorlinks,linkcolor=blue,citecolor=blue,urlcolor=blue]{hyperref}
\usepackage[normalem]{ulem}
\usepackage[all]{hypcap}
\usepackage[dvipsnames,usenames,table]{xcolor}
\usepackage{braket,mleftright}

\newcommand{\nocontentsline}[3]{}
\newcommand{\tocless}[2]{\bgroup\let\addcontentsline=\nocontentsline#1{#2}\egroup}

\newcommand{\bk}{{\bf k}}
\newcommand{\bK}{{\bf K}}
\newcommand{\bq}{{\bf q}}

\newcommand{\bz}{{\bf z}}

\newcommand{\bw}{{\bf w}}

\newcommand{\bn}{{\bf n}}

\newcommand{\btau}{\boldsymbol{\tau} }

\newcommand{\be}{\begin{equation}}
\newcommand{\ee}{\end{equation}}
\newcommand{\beg}{\begin{gather}}
\newcommand{\eeg}{\end{gather}}
\newcommand{\beq}{\begin{eqnarray}}
\newcommand{\eeq}{\end{eqnarray}}
\newcommand{\bea}{\begin{align}}
\newcommand{\eea}{\end{align}}
\newcommand{\beqq}{\begin{eqnarray*}}
\newcommand{\eeqq}{\end{eqnarray*}}

\newcommand{\up}{\uparrow}
\newcommand{\down}{\downarrow}

\newcommand{\ve}{{\varepsilon}}

\begin{document}

\title{Topological piezomagnetic effect in two-dimensional Dirac quadrupole altermagnets}

\author{H. Radhakrishnan}
\affiliation{Department of Physics, Drexel University, Philadelphia, PA 19104, USA}%
\affiliation{Department of Materials Science \& Engineering, Drexel University, Philadelphia, PA 19104, USA \looseness=-1}%

\author{Beryl Bell}
\affiliation{Department of Physics, Drexel University, Philadelphia, PA 19104, USA}%

\author{C. Ortix}
\affiliation{Dipartimento di Fisica ``E. R. Caianiello'', Universit\`a di Salerno, IT-84084 Fisciano (SA), Italy}%
\affiliation{CNR-SPIN, I-84084 Fisciano (Salerno), Italy, c/o Universit\'a di Salerno, I-84084 Fisciano (Salerno), Italy}

\author{J. W. F. Venderbos}
\affiliation{Department of Physics, Drexel University, Philadelphia, PA 19104, USA}%
\affiliation{Department of Materials Science \& Engineering, Drexel University, Philadelphia, PA 19104, USA \looseness=-1}%

\begin{abstract}
Altermagnets provide a natural platform for studying and exploiting piezomagnetism. In this paper, we introduce a class of insulating altermagnets in two dimensions (2D) referred to as Dirac quadrupole altermagnets, and show based on microscopic minimal models that the orbital piezomagnetic polarizability of such altermagnets has a topological contribution described by topological response theory. The essential low-energy electronic structure of Dirac quadrupole altermagnets can be understood from a gapless parent phase (i.e., the Dirac quadrupole semimetal), which has important implications for their response to external fields. Focusing on the strain-induced response, here we demonstrate that the topological piezomagnetic effect is a consequence of the way in which strain affects the Dirac points forming a quadrupole. We consider two microscopic models: a spinless two-band model describing a band inversion of $s$ and $d$ states, and a Lieb lattice model with collinear N\'eel order. The latter is a prototypical minimal model for altermagnetism in 2D and is realized in a number of recently proposed material compounds, which are discussed.  
\end{abstract}

\date{\today}

\maketitle

{\it Introduction.}---Topology is firmly established as one of the core organizing principles for our understanding of quantum phases of matter. One of the most compelling consequences of nontrivial topology in quantum materials are response properties fully determined by the topology of the ground state wave function. Examples include the quantized Hall response in the quantum Hall effect and Chern insulators~\cite{Laughlin:1981p5632,Halperin:1982p2185,Thouless:1982p405,Haldane:1988p2015}, and the quantized topological magnetoelectric effect of topological axion insulators in three dimensions~\cite{Qi:2008p195424,Essin:2009p146805,Li:2010p284,Zhang:2019p206401}. The extension and refinement of the classification of distinct topological phases have further deepened the understanding of the broader landscape of topological response phenomena originating from the structure of electronic wave functions. This extends to insulators as well as semimetals, and includes quantized and non-quantized responses~\cite{Montambaux:1990p11417,Balents:1996p2782,Haldane:2004p206602,Bernevig:2007p146804,Hughes:2011p245132,Goswami:2013p245107,Ramamurthy:2015p085105,Ramamurthy:2017p146602,deJuan:2017p15995}. Responses of primary interest and focus have been electromagnetic responses, thermal responses~\cite{Cooper:1997p2344,Wang:2011p014527,Ryu:2012p045104,Stone:2012p184503,Nomura:2012p026802,Bradlyn:2015p125303}, and geometric responses (e.g., responses to deformations and defects)~\cite{Avron:1995p697,Read:2009p045308,Hughes:2011p075502,Hughes:2013p025040,Parrikar:2014p105004,You:2016p085102,Sumiyoshi:2016p166601,Teo:2017p1,Gioia:2021p043067}, each highlighting the profound implications of nontrivial topology for observable material properties, as well as their compelling potential for applications. 

\begin{figure}
	\includegraphics[width=0.95\columnwidth]{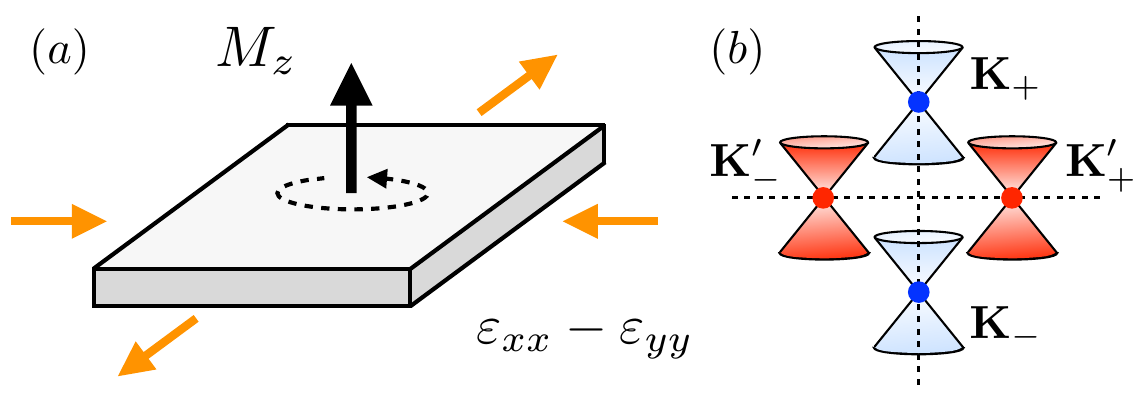}
	\caption{(a) Schematic depiction of the piezomagnetism in 2D altermagnets: an orbital magnetization $M_z$ develops in response to applied strain. (b) Configuration of Dirac points in a Dirac quadrupole semimetal, with positive and negative helicities color coded red and blue, respectively.}
	\label{fig1}
\end{figure}

In this paper, we focus on a distinct type of response property---the orbital piezomagnetic response---and predict a (non-quantized) topological piezomagnetic effect in a special class of magnetic insulators in two dimensions. The magnetic insulators in this class have two key characteristics: first, they are altermagnets, and second, they have a common parent state known as a ``Dirac quadrupole'' semimetal, i.e., a Dirac semimetal with four nodal crossings forming a quadrupole in reciprocal space (see Fig.~\ref{fig1})~\cite{Hirsbrunner:2024p041060}. We refer to this class of magnets as Dirac quadrupole altermagnets and show that the orbital piezomagnetic polarizability of Dirac quadrupole altermagnets has a (quasi)topological contribution captured by the topological response theory of the semimetallic parent state. 

Altermagnets are a recently proposed class of magnetic materials which combine properties of both ferromagnets (e.g., non-relativistic spin splitting) and antiferromagnets (e.g., compensated collinear magnetic order) \cite{Smejkal:2022p031042,Smejkal:2022p040501,Mazin:2022p040002,Turek:2022p094432,Mazin:2021pe2108924118,Song:2025p473,Jungwirth:2025p100162,Jungwirth:2026p837,Hayami:2019p123702,Bai:2023p216701,Yuan:2021p014409,Karube:2022p137201,Egorov:2021p2363,Guo:2023p100991,Naka:2019p4305,Yuan:2023pe2211966,Hayami:2020p144441,Gonzalez-Hernandez:2021p127701,Naka:2021p125114,Shao:2021p7061,Ma:2021p2846,Bose:2022p267}. This unusual combination of properties is enabled by the particular crystallographic environment of the magnetic atoms, and signals that crystal structure, magnetic order, and symmetry are interrelated in unconventional yet fundamental ways in altermagnets. One direct consequence of this interrelation is that altermagnets are natural candidates for piezomagnetic responses, i.e., the magnetization response to applied strain (sketched in Fig.~\ref{fig1})~\cite{Ma:2021p2846,McClarty:2024p176702,Aoyama:2024pL041402,Steward:2023p144418,Fernandes:2024p024404,Takahashi:2025p184408,Naka:2025p083702,Khodas:2025arXiv06}. This magnetization response generally has two contributions, the spin and the orbital contribution~\cite{Sorn:2025p245115}, and here we focus exclusively on the latter. The orbital magnetization is sensitive to the quantum geometric structure of the eigenstate wave functions, in particular the Berry curvature, and therefore carries a signature of band topology~\cite{Thonhauser:2005p137205,Xiao:2005p137204,Ceresoli:2006p024408,Mitscherling:2025arXiv12}. 

By considering two minimal models for Dirac quadrupole altermagnets with vanishing spin magnetization response, we show that the orbital magnetization response to strain is determined by the way in which strain distorts the Dirac quadrupole and creates a ``Dirac dipole''. The first model describes a generalized spinless altermagnetic state, hence emphasizing the orbital nature of topological piezomagnetism, and the second model describes a pure Lieb lattice altermagnet~\cite{Brekke:2023p224421,Antonenko:2025p096703}. The latter realizes a Dirac quadrupole semimetal in the non-relativistic limit~\cite{Antonenko:2025p096703} and has become a paradigmatic minimal model for the study of altermagnetism, in particular in two dimensions~\cite{Brekke:2023p224421,Antonenko:2025p096703,Yershov:2024p144421,Kaushal:2025p156502,Durrnagel:2025p036502,Fu:2025arXiv07}. Since a number of candidate altermagnets with Lieb lattice motif have recently been proposed~\cite{Ma:2021p2846,Wei:2025p024402,Chang:2025arXiv08} and investigated~\cite{Jiang:2025p754,Zhang:2025p760,Wei:2025p024402}, we discuss possible material realizations of the topological piezomagnetic effect. 

\begin{figure}
	\includegraphics[width=\linewidth]{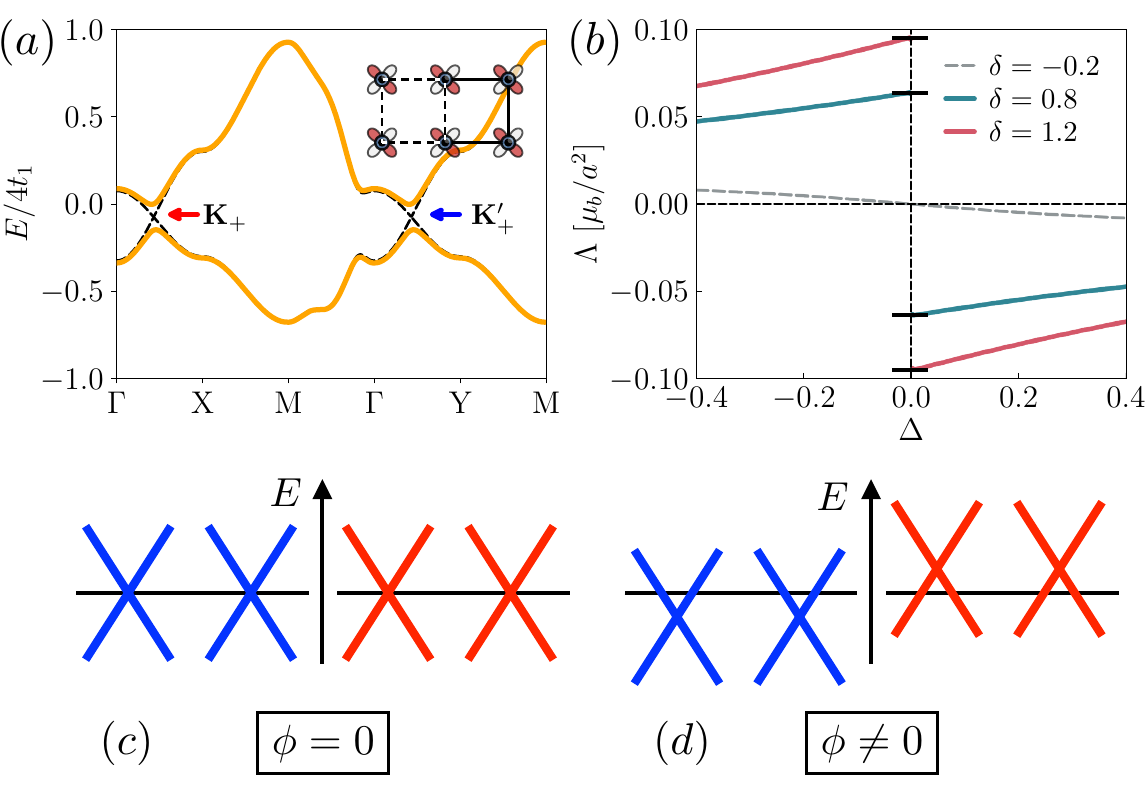}
	\caption{(a) Energy bands of the orbital altermagnet model defined in Eq.~\eqref{eq:Hk-orb} and schematically depicted in the inset (upper right corner). Here we have set $(t_0,t_d,\delta) = (0.25t_1,0.5t_1,0.8t_1)$; the orange and thin black curves correspond to $\Delta=0.3t_1$ and $\Delta=0.0$, respectively. The Dirac points and their valley labels are indicated. (b) Orbital piezomagnetic polarizability $\Lambda$ of the orbital altermagnet model, computed  using Eq.~\eqref{eq:Lambda-2band} and shown as a function of $\Delta$ (in units of $t_1$). Here we have used $(t_0,t_d) = (0.5t_1,0.5t_1)$. The red dashed lines correspond to $\pm (1/\pi) (t_0 \delta/t^2_1) $. (c) Without strain ($\phi=0$) all Dirac points occur at the same energy. (d) When strain is applied ($\phi \neq0$) nodes of opposite topological charge are shifted in opposite directions. }
	\label{fig2}
\end{figure}

{\it Orbital Dirac quadrupole altermagnet.}---We begin by introducing a spinless two-band model for a Dirac quadrupole altermagnet. Consider a simple square lattice with two orbitals per site, an $s$ orbital and a $d_{xy}$ orbital, as shown in the inset of Fig.~\ref{fig2}(a). The $s$ and $d$ states form the conduction and valence band, respectively, and we introduce the electron destruction operators to describe these states as $c_\bk = (s_\bk, d_\bk )^T$. With the tight-binding Hamiltonian defined as $H = \sum_\bk c^\dagger_\bk H_{0,\bk} c_\bk$, the Bloch Hamiltonian matrix $H_\bk $ takes the form
\be
H_{0,\bk} = \ve_\bk + t_{z,\bk}\tau^z +  t_{x,\bk}\tau^x + \Delta \tau^y, \label{eq:Hk-orb}
\ee
where $\btau = (\tau^x,\tau^y,\tau^z)$ is a set of Pauli matrices associated with the orbital degree of freedom. Here $\ve_\bk$ describes an orbital-independent dispersion given by $\ve_\bk = -t_0(\cos k_x + \cos k_y)$ and the functions $t_{z,\bk}$ and $t_{x,\bk}$ are given by $t_{z,\bk} = t_1(2-\cos k_x - \cos k_y) - \delta$ and $t_{x,\bk} = 4t_d \sin k_x \sin k_y$, respectively. These terms correspond to an intra-orbital nearest neighbor hopping amplitude $(t_0 \pm t_1)/2 $ for the $s$ and $d$ bands (with $t_1>t_0$) and a next-nearest neighbor interorbital hopping $t_d$. The form of $t_{x,\bk}$ is imposed by the symmetry of the $d_{xy}$ orbital, such that $H_{0,\bk}$ respects all symmetries of the square lattice when $\Delta = 0$. The parameter $\delta $ controls the relative energy of the $s$ and $d$ bands, and determines whether the bands are inverted ($\delta > 0$) or uninverted ($\delta < 0$). In the inverted regime and the fully symmetric limit $\Delta = 0$, $H_{0,\bk}$ realizes a Dirac quadrupole semimetal with Dirac points on the $k_x=0$ and $k_y=0$ lines, as sketched in Fig.~\ref{fig1}(b)~\cite{Hirsbrunner:2024p041060}. The coupling between the $s$ and $d_{xy}$ states described by $\Delta$ represents a symmetry-breaking term which gaps these Dirac points. It breaks both time-reversal ($\mathcal T$) symmetry and all spatial symmetries under which the $d_{xy}$ orbital is odd, in particular the fourfold rotation $\mathcal C_{4z}$. Crucially, the product of time-reversal and fourfold rotation, denoted $\mathcal T \mathcal C_{4z}$, is preserved, and in this sense Eq.~\eqref{eq:Hk-orb} describes a spinless system in the same symmetry class as a $d$-wave altermagnet~\cite{Smejkal:2022p031042,Chakraborty:2025arXiv09,Pan:2025arXiv10}. It then follows that from a symmetry perspective this ``orbital altermagnet'' supports the same orbital magnetization response to applied strain expected in spinful $d$-wave altermagnets.

The Dirac quadrupole structure is an essential characteristic of the orbital altermagnet, with important implications for the piezomagnetic response. To capture and examine these implications, we construct a continuum Dirac model, obtained from an expansion of the lattice Hamiltonian in the band inverted regime. Such a model consists of four Dirac points, or valleys, located at $\bK_\pm = (0,\pm k_D)$ and $\bK'_\pm = (\pm k_D, 0)$, as indicated in Fig.~\ref{fig1}(b), with corresponding Dirac Hamiltonians $\mathcal H_{\bK_\pm} $ and $\mathcal H_{\bK'_\pm} $. Here $k_D$ is the solutions of $\sin(k_D/2) =\sqrt{\delta/2t_1}$, as further detailed in Appendix~\ref{app:orbital-Dirac}, and the Dirac Hamiltonian at $\bK_\pm$ is given by
\be
\mathcal H_{\bK_\pm} = \pm (v_1 q_y \tau^z + v_2 q_x \tau^x ) + \Delta \tau^y, \label{eq:H-Dirac-orbital}
\ee
with Fermi velocities $v_{1,2}$ defined as $v_1 = \sqrt{\delta(2t_1-\delta)}$ and $v_2 = (4t_d/t_1) \sqrt{\delta(2t_1-\delta)}$. The Dirac Hamiltonian $\mathcal H_{\bK'_\pm}$ at valleys $\bK'_\pm$ is obtained from \eqref{eq:H-Dirac-orbital} by exchanging $q_x$ and $q_y$ (see Appendix~\ref{app:orbital-Dirac}). Equipped with this continuum Dirac model, we now turn to a calculation of the orbital piezomagnetic polarizability.

{\it Orbital piezomagnetic polarizability.}---The magnetization response to applied strain can be described by a general response equation of the form $M_i = \Lambda_{ijk} \varepsilon_{jk}$, where $M_i$ are the components of the magnetization and $\varepsilon_{jk}$ are the components of the strain tensor. The piezomagnetic tensor $\Lambda_{ijk}$ captures the linear piezomagnetic polarizability of the unstrained ground state and is thus constrained by the symmetries of the unstrained state~\cite{Khodas:2025arXiv06}.
In the case of the square lattice orbital altermagnet, the only nonzero components of the piezomagnetic tensor are $\Lambda_{zxx}$ and $\Lambda_{zyy} $, which are related as $\Lambda_{zxx}=-\Lambda_{zyy} \equiv \Lambda$. 

To compute the orbital piezomagnetic polarizability $\Lambda$ of the orbital altermagnet lattice model, we write the full Hamiltonian in the presence of strain as $H_\bk =  H_{0,\bk} + \phi W_\bk$, where $H_{0,\bk}$ is the unstrained Hamiltonian given by Eq.~\eqref{eq:Hk-orb}, $\phi$ denotes a strain field with $\varepsilon_{xx}-\varepsilon_{yy}$ symmetry, and $W_\bk$ describes the microscopic (and hence model-specific) implementation of strain. For purposes of generality it is convenient to choose a general parametrization of both $ H_{0,\bk}$ and $W_\bk$ as
\be
 H_{0,\bk} = \ve_\bk + \bn_\bk \cdot \btau, \quad W_\bk = \chi_\bk + \bw_\bk \cdot \btau, \label{eq:H0_W}
\ee
such that strain effects, in general, are specified by the functions $\chi_\bk$ and $\bw_\bk = (w_{x,\bk},w_{y,\bk},w_{z,\bk})$~\footnote{Basic symmetry requirements suggest that $w_{y,\bk} \equiv 0$.}, and $\bn_\bk = (t_{x,\bk},\Delta, t_{z,\bk})$. With this parametrization of the Hamiltonian and the strain perturbation the orbital piezomagnetic polarizability takes the form~\cite{Venderbos:arXiv2025-2}
\begin{multline}
\Lambda = 
- \frac{e}{\hbar} \int \frac{d^2\bk}{(2\pi)^2}\bigg\{ \chi_\bk \frac{\bn_\bk \cdot \partial_x \bn_\bk \times \partial_y \bn_\bk}{2|\bn_\bk|^3}  \\
 +\bigg[ \partial_x \varepsilon_\bk   \frac{\bw_\bk \cdot \bn_\bk \times \partial_y \bn_\bk}{2|\bn_\bk|^3} -  (x\leftrightarrow y) \bigg] \bigg\}. \label{eq:Lambda-2band}
\end{multline}
The polarizability is a sum of two terms, the ``geometric term'' and the ``interband term''~\cite{Venderbos:arXiv2025-2}, and the former has a straightforward interpretation: it is the integral over the Berry curvature of the occupied valence band multiplied by $\chi_\bk$, the function encoding the strain coupling. This signals that $\Lambda$ is sensitive to the topology of the occupied band. To compute $\Lambda$ for the orbital altermagnet model, we choose a strain coupling given by $\chi_\bk=-t_0(\cos k_x - \cos k_y)$ and $\bw_\bk=0$. This corresponds to a change of the hopping $t_0 \rightarrow t_0(1\pm \phi)$ along the $x/y$ direction, and since $\bw_\bk=0$, only the first term (i.e., the `geometric'' term) in \eqref{eq:Lambda-2band} contributes. More general strain couplings are discussed in Appendix~\ref{app:orbital-strain}. 

In Fig.~\ref{fig2}(b) we show $\Lambda$ as a function of $\Delta$ for different values of $\delta$. A striking feature observed in Fig.~\ref{fig2}(b) is the finite limiting value of $\Lambda$ as $\Delta \rightarrow 0$ when $\delta>0$ (i.e., in the band-inverted regime), giving rise to a discontinuity at $\Delta = 0$. This is counterintuitive, since $\Delta \rightarrow 0$ corresponds to a limit in which $\mathcal T$, which forbids a piezomagnetic response, is restored. The nonzero polarizability in the limit of vanishing altermagnetic order $\Delta$ is a direct consequence of the Dirac quadrupole semimetallic parent state of the orbital altermagnet, and a manifestation of the topological response theory governing topological semimetals in two dimensions~\cite{Ramamurthy:2015p085105}.

{\it Topological response theory.}---The discontinuity observed in Fig.~\ref{fig2}(b) can be understood from a calculation of $\Lambda$ based on the continuum Dirac model given by Eq.~\eqref{eq:H-Dirac-orbital} (i.e., the Dirac Hamiltonians $\mathcal H_{\bK_\pm}$ and $\mathcal H_{\bK'_\pm}$). Application of Eq.~\eqref{eq:Lambda-2band} to each Dirac valley yields the integrated Berry curvature multiplied by $\chi_\bk$ evaluated at the valley momentum $\bK_\pm$ or $\bK'_\pm$ (see Appendix~\ref{app:orbital-Dirac} for details). Defining $\chi_{\bK_\pm}=-\chi_{\bK'_\pm}\equiv w^D_0$, with $w^D_0 = t_0\delta/t_1$, we obtain $\Lambda = -(e/\pi \hbar)w^D_0 \text{sgn}(\Delta)$, which, as indicated in Fig.~\ref{fig2}(b), exactly matches the discontinuity found in the lattice calculation. This result can be understood on a more fundamental level from the topological response theory of Dirac semimetals in two dimensions, which predicts an orbital magnetization response when the Dirac points are shifted in energy~\cite{Ramamurthy:2015p085105}. Specifically, it predicts an orbital magnetization given by $M_z = -(e/2\pi)b_0 \text{sgn}(\Delta)$, where $\hbar b_0 = \sum_n \ve^D_n Q_n$ is the energy ``dipole'' of a configuration of Dirac points with energy $\ve^D_n$ and topological charge $Q_n$~\cite{Ramamurthy:2015p085105}. In the orbital altermagnet model such an energy dipole is created by strain, which shifts the Dirac points with positive (negative) charge up (down) in energy by an amount $\phi w^D_0$, as sketched in Fig.~\ref{fig2}(c-d). Hence, $b_0 = 2\phi w^D_0/\hbar$ such that the topological response equation yields $\partial M_z /\partial \phi = -(e/\pi \hbar)w^D_0 \text{sgn}(\Delta)$, in full agreement with continuum model calculation based on Eq.~\eqref{eq:Lambda-2band}. Our central result is therefore that the orbital piezomagnetic polarizability of a Dirac quadrupole altermagnet has a topological contribution derived from a general topological response action. 

\begin{figure}
	\includegraphics[width=\columnwidth]{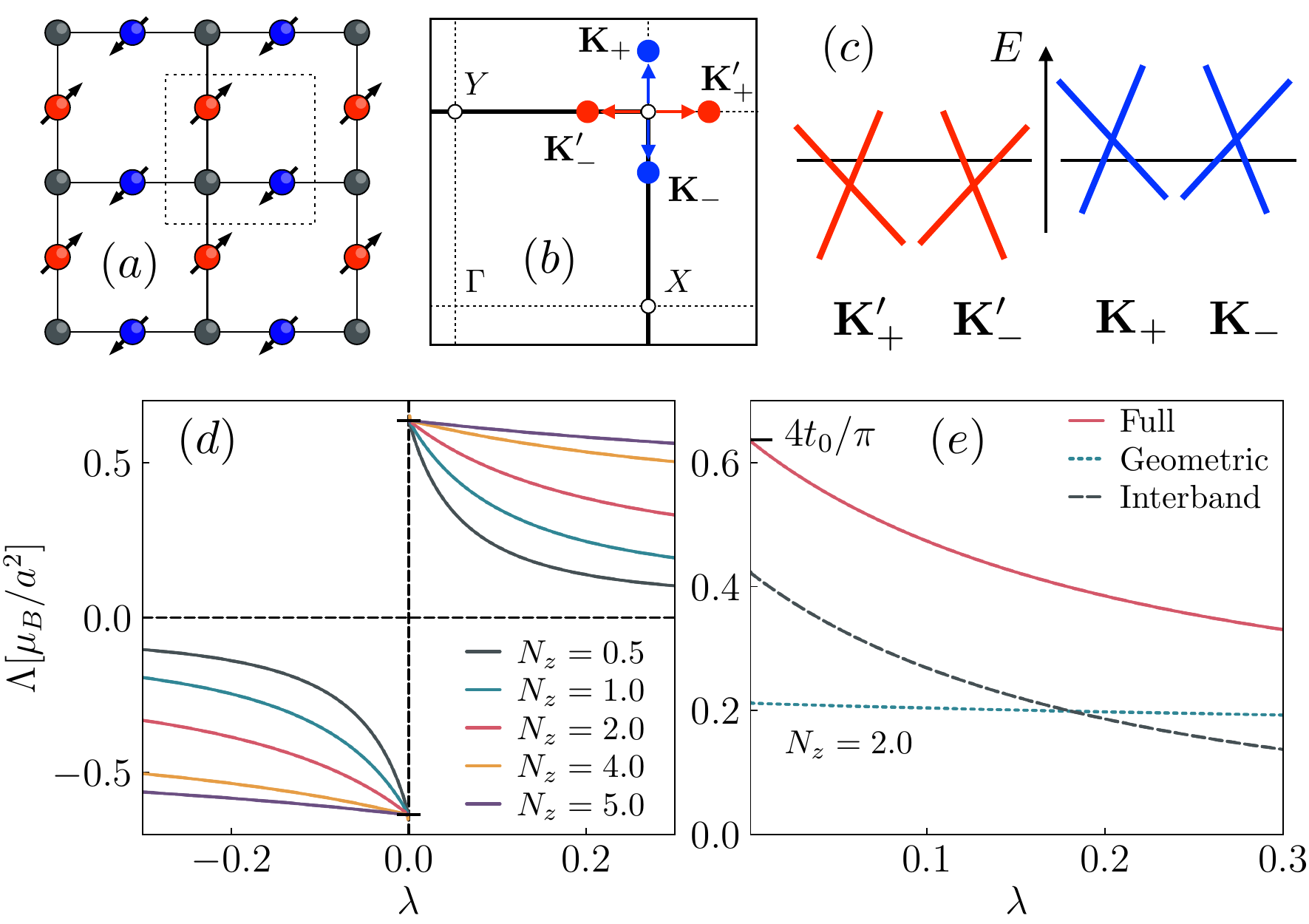}
	\caption{(a) Lieb lattice altermagnet model. The non-magnetic sites are shown in grey and the magnetic sites are shown in red and blue ($A$ and $B$ sublattice, respectively), indicating opposite alignment of spin moments along the $\hat z$ direction. (b) Location of the Dirac points on the BZ boundary in the topological regime ($|N_z|<4t_d$). Valleys labeled red and blue correspond to the $\sigma=\up$ and $\sigma=\down$ sectors, respectively (assuming $N_z>0$, see Appendix~\ref{app:Lieb}). (c) Effect of strain on the (tilted) Dirac points. The Dirac points of the valleys $\mathbf{K}_\pm$ and $\mathbf{K}'_\pm$ are shifted in energy in opposite directions. (d) Orbital piezomagnetic polarizability as a function of $\lambda$ for different $N_z$ (both in units of $t_1$), calculated using  a generalization of Eq.~\eqref{eq:Lambda-2band} (see Appendix~\ref{app:piezomagnet-Lieb}). Here we have set $(t_0,t_d) = (0.5t_1,1.5t_1)$. (e) Same as in (d), showing contributions from the geometric and interband terms (for $N_z=2.0t_1$).}
	\label{fig3}
\end{figure}

{\it Lieb lattice Dirac quadrupole altermagnet.}---We now turn to the spinful Lieb lattice model shown in Fig.~\ref{fig3}(a). The Lieb lattice consists of a nonmagnetic site (indicated in gray) at the corner of each square plaquette, and two magnetic sites at the center of the edges (labeled $A$ and $B$ and indicated in red and blue). The two magnetic sites are exchanged by a $\mathcal C_{4z}$ rotation, but not by inversion or a translation, and as a result, collinear magnetic order with anti-aligned moments on the $A$ and $B$ sublattices leads to an altermagnetic state. In what follows we assume that the moments order along the $\hat z$ axis, i.e., perpendicular to the plane, in which case $\mathcal T \mathcal C_{4z}$ symmetry forbids a net magnetization in the presence of spin-orbit coupling~\cite{Antonenko:2025p096703}. Altermagnets of this kind are referred to as `pure' altermagnets, as opposed to `mixed' altermagnets, which permit weak ferromagnetism when spin-orbit coupling is included~\cite{Fernandes:2024p024404}. 

The Lieb lattice altermagnet with N\'eel vector along the $\hat z$ axis preserves a horizontal mirror symmetry $\mathcal M_z$, which commutes with the Hamiltonian at each $\bk$ and therefore block diagonalizes the Hamiltonian into disconnected mirror eigenvalue sectors. Since these eigenvalues physically correspond to spin, the Hamiltonian is manifestly diagonal in spin space and can be considered in each spin sector ($\sigma$) separately. Specifically, we can write the Hamiltonian of the pure Lieb lattice altermagnet as
\begin{equation}
    H_{0,\mathbf{k}}^\sigma = \ve_\mathbf{k}+\mathbf{n}_\mathbf{k}^\sigma \cdot \mathbf{\tau}, \label{eq:H_0-Lieb}
\end{equation}
where the Pauli matrices $\btau$ now correspond to sublattice degree of freedom, i.e., the tight-binding spinor $c_\bk = (c_{\bk A}, c_{\bk B})^T$, and $\sigma=\up,\down$ corresponds to spin. Here, the dispersion $\ve_\bk =-2t_0(\cos{k_x}+\cos{k_y})$ is proportional to the identity and describes an intra-sublattice next-nearest neighbor hopping. The vectorial function $\bn^\sigma_\bk$ is written as $\bn^\sigma_\bk = (t_{x,\bk}, \sigma \lambda_{z,\bk}, t_{z,\bk}+\sigma N_z)$, where $t_{x,\bk} = -4t_1\cos(k_x/2)\cos(k_y/2)$ derives from an intra-sublattice nearest neighbor hopping, and $t_{z,\bk} =-2t_d(\cos{k_x}-\cos{k_y})$ describes an anisotropic next-nearest neighbor hopping allowed by the anisotropic local crystal environments seen by the $A$ and $B$ sites. Furthermore, $N_z$ is the $\hat z$-component of N\'eel vector (i.e., intra-unit cell staggered magnetization) and $\lambda_{z,\bk} = 4\lambda \sin(k_x/2)\sin(k_y/2) $ captures a spin-orbit coupling term of Kane-Mele type~\cite{Kane:2005p226801,Weeks:2010p085310}. 

A key feature of the Lieb lattice model is the existence of linear Dirac crossings on the Brillouin zone (BZ) boundary in the limit of vanishing spin-orbit coupling ($\lambda=0$), and provided $|N_z|/4t_d<1$~\cite{Antonenko:2025p096703}. One pair of Dirac points is realized in each spin sector, such that the two pairs of Dirac points are related by $\mathcal T \mathcal C_{4z}$, i.e., the product of time-reversal and fourfold rotation---an example of crystal symmetry-paired spin-valley locking~\cite{Ma:2021p2846}. The location of the Dirac points is sketched in Fig.~\ref{fig3}(b), with red and blue corresponding to opposite spin ($\sigma=\up$ and $\sigma=\down$, respectively, when $N_z>0$). When spin-orbit coupling is included, the Dirac points are gapped, and the Berry curvature in the spin-projected valleys $\bK_\pm$ and $\bK'_\mp$ has opposite sign. This is consistent with $\mathcal T \mathcal C_{4z}$ symmetry (and is reflected in the color coding in Fig.~\ref{fig3}) and shows that the Lieb lattice model realizes a Dirac quadrupole altermagnet similar to the orbital altermagnet model defined in Eq.~\eqref{eq:Hk-orb}. Due to the spin-valley locking the Lieb lattice altermagnet further realizes a mirror Chern insulator at half-filling. Here we are focused on this half-filled insulator, for which the spin piezomagnetic polarizability vanishes~\cite{Takahashi:2025p184408}.

To examine the consequences of the Lieb lattice Dirac quadrupole structure, in particular for the piezomagnetic polarizability, we construct a linearized continuum model by expanding around the Dirac points. The Dirac points are located at $\bK_\pm = (\pi,\pi\pm k_D)$ and $\bK'_\pm = (\pi\pm k_D,\pi)$, with $k_D \in (0,\pi)$ given by the solution to the equation $\sin(k_D/2)=\sqrt{|N_z|/4t_d}$, and thus set by the magnitude of $N_z$ (see Appendix~\ref{app:Lieb}). An expansion around these points yields the Dirac Hamiltonians
\begin{align}
\mathcal{H}_{\mathbf{K}_\pm} &= \mp v_0q_y \mp( v_1q_x\tau^x-v_2q_y\tau^z)- \zeta m\tau^y, \label{eq:Lieb-Dirac-K} \\
\mathcal{H}_{\mathbf{K}'_\pm} &= \mp v_0q_x \mp(v_1q_y\tau^x+v_2q_x\tau^z)+\zeta m\tau^y, \label{eq:Lieb-Dirac-K'}
\end{align}
where $v_1$ and $v_2$ are two Dirac velocities given by $v_1 = 2t_1\sqrt{|N_z|/4t_d}$ and $v_2 = 4t_d\sqrt{(|N_z|/4t_d)(1-|N_z|/4t_d)}$, and $m = 4\lambda \sqrt{1-|N_z|/4t_d}$ is a Dirac mass. We have further defined $\zeta \equiv \text{sgn}(N_z)$. The first term in Eqs.~\eqref{eq:Lieb-Dirac-K} and \eqref{eq:Lieb-Dirac-K'} describes a tilt of the Dirac nodes, as sketched in Fig.~\ref{fig3}(c), with tilt velocity $v_0 =  4t_0\sqrt{(|N_z|/4t_d)(1-|N_z|/4t_d)}$. Note that the Dirac mass $m$ is indeed proportional to $\lambda$, the strength of spin-orbit coupling, such that in the limit $\lambda \rightarrow 0$ the Dirac points become gapless.

Next, we include the effect of strain in the Lieb lattice model. As in the orbital altermagnet model, in order to generate a magnetization response the applied strain must be of $\varepsilon_{xx}-\varepsilon_{yy}$ symmetry. The general form of the strain coupling is chosen as in Eq.~\eqref{eq:H0_W}, such that the full Lieb lattice Hamiltonian in the presence of strain is $H_{\mathbf{k}}^\sigma=H_{0,\mathbf{k}}^\sigma+\phi W_\bk$. The specific coupling to strain is chosen as $\chi_\bk  = -2t_0(\cos k_x-\cos k_y)$ and $w_{z,\bk}  = -2t_d(\cos k_x+\cos k_y)$, with $w_{x,\bk}=w_{y,\bk}=0 $ (see Ref.~\onlinecite{Takahashi:2025p184408}). On the $A$ sublattice, this corresponds to a next-nearest neighbor hopping amplitude $(t_0 + t_d)(1+\phi) $ in the $\hat x$ direction, and $(t_0 - t_d)(1-\phi) $ in the $\hat y$ direction; on the $B$ sublattice the sign of $t_d$ is reversed~\cite{Venderbos:arXiv2025-2}. Within the linearized continuum model the effect of strain is determined by evaluating $\chi_\bk$ and $ w_{z,\bk}$ at $\bK_\pm$ and $\bK'_\pm$. We define $\chi_{\bK_\pm} = -\chi_{\bK'_\pm} \equiv w_0^D$ and $w_{z,\bK_\pm} = w_{z,\bK'_\pm} \equiv w_z^D$, such that the strain coupling $W_\bk$ enters the continuum model Hamiltonian at $\bK_\pm$ and $\bK'_\pm$ as $\mathcal W_{\bK_\pm} = w^D_0 + w^D_z \tau^z$ and $\mathcal W_{\bK'_\pm} = -w^D_0 + w^D_z \tau^z$, respectively. Here $w^D_0$ corresponds to a shift in energy, which is opposite for the valleys of opposite Berry curvature, whereas $w^D_z$ corresponds to a shift of the Dirac points in momentum. 

{\it Lieb lattice piezomagnetic polarizability.}---The expression for the orbital piezomagnetic polarizability applicable to the Lieb lattice model is a sum over contributions from each spin sector, with the contribution from each spin sector given by the two-band formula of Eq.~\eqref{eq:Lambda-2band} (see Appendix \ref{app:piezomagnet-Lieb}). In Fig.~\ref{fig3}(d) we show the piezomagnetic polarizability of the Lieb lattice model as a function of $\lambda$, the spin-orbit interaction, for different values of $N_z$. It exhibits a discontinuity at $\lambda=0$ similar to Fig.~\ref{fig2}(b), which is at odds with the expectation that an orbital piezomagnetic effect is forbidden when full spin-rotation invariance is restored in the limit $\lambda \rightarrow 0$. As was the case for the orbital altermagnet, the finite limiting value of $\Lambda$ corresponds to a topological contribution to the polarizability and originates from the Dirac quadrupole parent state. The topological contribution can be obtained from a calculation of $\Lambda$ within the continuum Dirac model given by Eqs.~\eqref{eq:Lieb-Dirac-K} and \eqref{eq:Lieb-Dirac-K'}, combined with the strain terms $\mathcal W_{\bK_\pm}$ and $\mathcal W_{\bK'_\pm}$. As detailed in Appendix~\ref{app:piezomagnet-Lieb}, this yields $\Lambda \simeq  \text{sgn}(\lambda N_z) (e/\pi\hbar) \left(w^D_0 + v_0w^D_z/v_2 \right) = \text{sgn}(\lambda N_z) (4t_0 e/\pi\hbar)$, and reveals that both the geometric term and the interband term give rise to a topological contribution (proportional to $w^D_0$ and $w^D_z$, respectively). This is evidenced by Fig.~\ref{fig3}(e), which separately shows the contributions of the geometric and interband terms to $\Lambda$.  The topological contribution from the geometric term can be understood as before from the way applied strain shifts the Lieb lattice Dirac points in energy and creates a Dirac dipole. This is shown in Fig.~\ref{fig3}(c). The contribution from the interband term is a consequence of its interpretation in terms of a generalized band curvature~\cite{Venderbos:arXiv2025-2}, which is specific to two-band systems. In the case of two-band Dirac models this generalized curvature is proportional to the standard Berry curvature, and yields a topological contribution to $\Lambda$ provided the Dirac cones are tilted, as sketched in Fig.~\ref{fig3}(c) (see Appendix \ref{app:piezomagnet-Lieb}). 

{\it Discussion and conclusion.}---The central result of this paper is the prediction of a topological orbital piezomagnetic effect in 2D altermagnets described by the topological response theory of Dirac semimetals. According to this response theory, the orbital magnetization $M_z$ of a Dirac semimetal in 2D is proportional to the time component of the Dirac point energy-momentum dipole moment. Since the latter is generically not quantized, the response theory of topological semimetals describes non-quantized responses, in contrast to the quantized responses of topological insulators. 

The topological orbital piezomagnetic effect occurs in a specific family of altermagnets here referred to as Dirac quadrupole altermagnets. These can be described by a continuum theory of four gapped Dirac points which realize a momentum quadrupole moment protected by $\mathcal T\mathcal C_{4z}$ symmetry, and provide a premier venue for studying manifestations of nontrivial topology in magnetically ordered systems. 


Here we have examined two minimal lattice models of Dirac quadrupole altermagnets. The two-band orbital altermagnet model, in which the Dirac quadrupole arises as a result of a band inversion, clearly exposes the orbital nature of the effect---since spin is absent---and provides the simplest nontrivial model for studying the interplay of altermagnetism and topology. The Lieb lattice model has not only garnered broad recognition as a paradigmatic minimal model for studying a variety of different phenomena related to altermagnetism, but also establishes a direct connection to materials of active current interest. A number of materials with Lieb lattice motif have been identified or proposed~\cite{Chang:2025arXiv08}, providing a realistic platform for experimentally probing orbital piezomagnetism. A promising family of materials includes the layered bulk compounds V$_2$Se$_2$O and V$_2$Te$_2$O, as well their Rb or K-intercalated variants~\cite{Lin:2018p075132,Ma:2021p2846,Li:2024p222404,Jiang:2025p754,Zhang:2025p760}, which realize the Lieb lattice in monolayer form. The correlated insulator La$_2$O$_3$Mn$_2$Se$_2$ is another confirmed altermagnet in which the MnO$_2$ planes form a Lieb lattice~\cite{Wei:2025p024402}. A promising proposed metallic compound is Sr$_2$CrO$_2$Cr$_2$OAs$_2$, which is predicted to exhibit a large altermagnetic magnon splitting~\cite{Chang:2025arXiv08}.

{\it Acknowledgements.}---We gratefully acknowledge insightful conversations with Rafael Fernandes, Daniel Agterberg, and Taylor Hughes. B.B., H.R., and J.W.F.V. were supported by the U.S. Department of Energy under Award No. DE-SC0025632. C.O. acknowledges partial support by the Italian Ministry of Foreign Affairs and International Cooperation PGR12351 (ULTRAQMAT) and from PNRR MUR Project No. PE0000023-NQSTI (TOPQIN).

\bibliography{altermagnet,topo_response}

\appendix

\section{Linearized continuum model for the orbital altermagnet}
\label{app:orbital-Dirac}

This Appendix collects further details on the construction of a continuum Dirac model for the orbital altermagnet. First, we perform an expansion of the lattice model around the Dirac points in the absence of strain. Then we determine how the strain term considered in the main text enters the continuum Dirac model. In Appendix~\ref{app:orbital-strain} we develop a more general theory of strain coupling.

{\it Construction of continuum Dirac model without strain.} The first step in the derivation of a continuum theory is to determine the location of the Dirac points in the limit $\Delta=0$. The momenta where the Dirac points are located are the solutions of the equation $t_{z,\bk} =t_1(2-\cos k_x-\cos k_y )- \delta =0$, subject to the constraint $k_x=0$ or $k_y=0$. Consider first the case $k_x=0$. The equation we seek to solve reduces to $t_1(1-\cos k_y)- \delta = 2t_1 \sin^2(k_y/2) -\delta=0$, from which it directly follows that 
\be
\sin( k_y/2) = \pm \sqrt{\delta/2t_1}.
\ee
These two solutions correspond to the valleys labeled $\bK_\pm = (0,\pm k_D)$ in Fig.~\ref{fig1}(b), where we have defined $k_D$ as the solution of $\sin( k_y/2) = \sqrt{\delta/2t_1}$ in the interval $k_y \in (0,\pi)$. It then immediately follows from symmetry arguments that the Dirac valleys on the $k_x$ axis are given by $\bK'_\pm = (\pm k_D,0)$.

The next step is to expand the lattice Hamiltonian to linear order in small momentum $\bq$ relative to the Dirac point momenta. Consider the valleys $\bK_\pm$. Expanding the functions $t_{z,\bk}$ and $t_{x,\bk}$ yields
\begin{align}
t_{z,\bK_\pm +\bq} &\approx \pm t_1 \sin k_D q_y, \\
t_{x,\bK_\pm +\bq} &\approx \pm 4t_d \sin k_D q_x,
\end{align}
where we can write $\sin k_D = 2\sin( k_D/2)\cos( k_D/2) $. Now using that $\sin( k_D/2) = \sqrt{\delta/2t_1}$ one finds $\sin k_D =  2\sqrt{(\delta/2t_1) (1-\delta/2t_1)}$. It is then natural to define the two Dirac velocities $v_1$ and $v_2$ as
\begin{align}
v_1 &\equiv t_1 \sin k_D = \sqrt{\delta(2t_1-\delta)},\label{app:v_1-orb} \\
v_2 &\equiv 4t_d \sin k_D  = \frac{4t_d}{t_1} \sqrt{\delta(2t_1-\delta)}. \label{app:v_2-orb}
\end{align}
Next, we expand the function $\ve_{\bk}$, which multiplies the identity. We find
\begin{align}
\ve_{\bK_\pm +\bq}& \approx -2t_0\cos^2\frac{k_D}{2}  \pm t_0 \sin k_D q_y, \label{app:eps-orb-expand-1} \\
\ve_{\bK'_\pm +\bq}& \approx -2t_0\cos^2\frac{k_D}{2}  \pm t_0 \sin k_D q_x, \label{app:eps-orb-expand-2}
\end{align}
and this motivates defining the velocity $v_0$ as
\be
v_0 \equiv t_0 \sin k_D = \frac{t_0}{t_1} \sqrt{\delta(2t_1-\delta)} \label{app:v_0-orb}
\ee
and the Dirac point energy $\ve_D$ as $\ve_D \equiv -2t_0\cos^2(k_D/2) = -t_0(2- \delta/t_1)$. The Dirac Hamiltonian for the valleys $\bK_\pm$ is then given by (ignoring the constant term $\ve_D$)
\be
\mathcal H_{\bK_\pm} =\pm v_0 q_y \pm (v_1 q_y \tau^z + v_2 q_x \tau^x ) + \Delta \tau^y. \label{app:Dirac-H-orb}
\ee
The Dirac Hamiltonian for the valleys $\bK'_\pm$ is obtained directly by exchanging $q_x$ and $q_y$. These Dirac Hamiltonians describe a gapped, anisotropic, and tilted (velocity $v_0$) Dirac cone. The gap is due to the mass term $\Delta$ and the anisotropy is a result of the generically different velocities $v_1$ and $v_2$. The tilt is caused by $\pm v_0 q_y$.
In the main text we have ignored this tilt due to $v_0$, since it does not enter the expression for the piezomagnetic polarizability when the strain coupling is chosen as $W_\bk = -t_0(\cos k_x - \cos k_y)$.

The continuum model thus consists of four gapped Dirac Hamiltonians for each of the four valleys. We note that, in the notation introduced in Eq.~\eqref{eq:H0_W} of the main text, the Dirac Hamiltonian for the valleys $\bK_\pm$ can be expressed as
\be
\mathcal H_{\bK_\pm} = \ve^{\bK_\pm}_\bq + \bn^{\bK_\pm}_\bq \cdot \btau, \label{app:Dirac-Kpm}
\ee
with
\be
\ve^{\bK_\pm}_\bq = \pm v_0 q_y, \quad \bn^{\bK_\pm}_\bq = (\pm v_2 q_x ,\Delta, \pm v_1 q_y  ). \label{app:eps-n-Kpm}
\ee
A similar expression holds for the valleys $\bK'_\pm$.

{\it Berry curvature.} Using this continuum model expansion, it is straightforward to compute the Berry curvature $\Omega_{xy}$ associated with the occupied valence band for each valley. We find the Berry curvatures of the different valleys as
\begin{align}
\Omega^{\bK_\pm}_{xy}& = -\frac{\Delta v_1 v_2}{2(\Delta^2 +v_1^2q_y^2+v_2^2q_x^2)^{3/2}}, \label{app:Omega_K-2B} \\
\Omega^{\bK'_\pm}_{xy} &= \frac{\Delta v_1 v_2}{2(\Delta^2 +v_1^2q_x^2+v_2^2q_y^2)^{3/2}} , \label{app:Omega_K'-2B}
\end{align}
which shows that the continuum model indeed gives rise to a Dirac quadrupole structure. The integrated Berry curvature yields
\be
\int \frac{d^2\bq}{2\pi} \Omega^{\bK_\pm}_{xy}= -\frac12 \text{sgn}(\Delta), \; \int \frac{d^2\bq}{2\pi}  \Omega^{\bK'_\pm}_{xy}= \frac12 \text{sgn}(\Delta), \label{app:top-Dirac}
\ee
as expected for Dirac points in 2D.

\section{Generalized strain coupling in the orbital altermagnet}
\label{app:orbital-strain}

In the main text we have considered a particular choice of strain coupling, namely we have chosen $W_\bk $ proportional to the identity: $W_\bk  = \chi_\bk = -t_0(\cos k_x - \cos k_y)$. In this Appendix we present the general form of $W_\bk$ allowed by symmetry, i.e., a form for which $\bw_\bk$ is nonzero, and discuss how it affects the calculation of the orbital piezomagnetic polarizability. 
 
{\it Generalization of strain coupling.} The orbital altermagnet model describes a band inversion of an $s$ and a $d$ band. In general, strain affects each band separately and leads to independent anisotropic dispersions of the $s$ and $d$ bands. This is captured by a strain perturbation $W_\bk$ given by
\be
\chi_\bk = -t_{a}(c_x - c_y), \quad w_{z,\bk} = -t_{b}(c_x - c_y) \label{app:strain-orb-gen}
\ee
Here we have abbreviated $c_j \equiv \cos k_j$ and $t_a$ and $t_b$ are introduced as independent hopping amplitudes (different from $t_0$ and $t_1$) describing the effect of strain. To see how this relates to anisotropic hopping within the $s$ and a $d$ bands, we first note that nearest neighbor hopping amplitudes between the $s$ and $d$ states are given by $t_s = (t_0 + t_1)/2$ and $t_d = (t_0 - t_1)/2$, respectively. In the presence of strain, described by the (dimensionless) strain field $\phi$, these hoppings change as
\begin{align}
t_s = \frac12(t_0 + t_1) & \rightarrow t_s(1 \pm \eta_s \phi) \\
t_d = \frac12(t_0 - t_1) & \rightarrow t_d(1 \pm \eta_d \phi)
\end{align}
where $\pm$ corresponds to the hopping along the $x/y$ direction, in accordance with an applied strain of $\ve_{xx}-\ve_{yy}$ symmetry. Here $\eta_s$ and $\eta_d$ are dimensionless parameters describing the generically different response of the $s$ and $d$ band hoppings to applied strain. It is then useful to define
\be
\eta_\pm  \equiv \frac12(\eta_s \pm \eta_d), \label{app:eta_pm-def}
\ee
such that the hoppings $t_a$ and $t_b$ introduced in Eq.~\eqref{app:strain-orb-gen} can be expressed in terms of $t_0 $ and $t_1$ as
\begin{align}
t_a & = \eta_+ t_0 + \eta_- t_1, \label{app:t_a}\\
t_b  & = \eta_+ t_1 + \eta_- t_0. \label{app:t_b}
\end{align}
Alternatively, one may express $\eta_\pm$ in terms of the hopping amplitudes $t_{0,1}$ and $t_{a,b}$ as
\be
\eta_+ = \frac{t_0 t_a - t_1 t_b}{t^2_0-t^2_1}, \quad \eta_- = \frac{t_0 t_b - t_1 t_a}{t^2_0-t^2_1}. \label{app:eta_pm}
\ee
Based on this discussion, the strain perturbation of Eq.~\eqref{app:strain-orb-gen} can be expressed in terms of $t_{0,1}$ and $\eta_\pm$ as
\begin{align}
\chi_\bk & = -(\eta_+ t_0 + \eta_- t_1)(\cos k_x - \cos k_y), \label{app:chi_k} \\
w_{z,\bk}  & = -(\eta_+ t_1 + \eta_- t_0)(\cos k_x - \cos k_y) \label{app:wz_k},
\end{align}
which represents the most general way in which strain can modify the hopping amplitudes of the $s$ and $d$ bands. 

{\it Further generalization.} As defined in Eq.~\eqref{eq:H0_W}, the most general form of the strain coupling in the two-band orbital altermagnet model is $W_\bk  = \chi_\bk + \bw_\bk \cdot \btau$. Above we determined the form of $\chi_\bk$ and $w_{z,\bk}$, which correspond to intra-band strain terms. Here we briefly comment on additional symmetry-allowed strain interband couplings that enter via $\bw_\bk$, in particular $w_{x,\bk}$. The requirement is that all terms in $W_\bk$ have the symmetry of the applied strain, which is of $d_{x^2-y^2}$ type. An additional term which satisfies this requirement is given by
\be
w_{x,\bk} = 8t_g \sin k_x \sin k_y (\cos k_x - \cos k_y). \label{app:wx}
\ee
This term represents an anisotropic fourth-nearest neighbor interorbital hopping (denoted $t_g$). One may recognize that the function $w_{x,\bk}$ takes the form of a square lattice $g$-wave form factor. This yields the required $d_{x^2-y^2}$ symmetry since the interorbital hopping is between an $s$-wave orbital and a $d_{xy}$ orbital. It is straightforward to determine that this additional term does not change the location of the Dirac points, either in energy or in momentum, and therefore does not affect the topological responses discussed in the main text.

\section{Piezomagnetic polarizability of the orbital altermagnet}
\label{app:orbital-piezomagnet}

In this Appendix we describe in more detail how to calculate the orbital piezomagnetic polarozability given by Eq.~\eqref{eq:Lambda-2band} of the main text. We first briefly comment on the full lattice model calculation. Then we determine how the strain coupling enters in the continuum model formulation. This serves as a basis for calculating the piezomagnetic polarizability within the continuum Dirac model. 

{\it Lattice model calculation of polarizability.} In the lattice model calculation of the piezomagnetic polarizability discussed in the main text, we evaluate Eq.~\eqref{eq:Lambda-2band} numerically using the full lattice tight-binding model. Choosing the strain coupling as $W_\bk = \chi_\bk+ w_{z,\bk}\tau^z$, with $\chi_\bk$ and $w_{z,\bk}$ given by Eqs.~\eqref{app:chi_k} and \eqref{app:wz_k}, the integrand of Eq.~\eqref{eq:Lambda-2band} may be obtained analytically, and we find
\be
\Lambda = \frac{2e\eta_- (t^2_0-t^2_1)}{\hbar} \int \frac{d^2\bk}{(2\pi)^2}\frac{t_d\Delta(c_x-c_y)^2(1+c_xc_y)}{(\Delta^2+t^2_{x,\bk}+t^2_{z,\bk})^{3/2}}. \label{app:Lambda-2b-lattice}
\ee
Here we have abbreviated $c_j \equiv \cos k_j$. Note that, based on Eq.~\eqref{app:eta_pm}, we may also write $\eta_- (t^2_0-t^2_1) = t_0 t_b - t_1 t_a$. The results shown in Fig.~\ref{fig2}(b) are obtained from numerically evaluating Eq.~\eqref{app:Lambda-2b-lattice} for the case $t_b=0$ and $t_a=t_0$. 


{\it Strain coupling in the continuum model.} To compute the piezomagnetic polarizability within the continuum model, we require an expansion of the strain term $W_\bk $. To obtain such an expansion we proceed as in Appendix~\ref{app:orbital-Dirac}. The lowest order contribution to an expansion of the function $\chi_\bk$ is a constant term (i.e., the value of $\chi_\bk$ at $\bK_\pm$ and $\bK'_\pm$), and we find
\begin{align}
\chi_{\bK_\pm } & = -(\eta_+ t_0 + \eta_- t_1)(1-\cos k_D ) ,\\
& =  - (\eta_+ t_0 + \eta_- t_1)\delta/ t_1  = - \chi_{\bK'_\pm } .
\end{align}
Similarly, the lowest order contribution to $w_{z,\bk}$ is found as
\begin{align}
w_{z,\bK_\pm } & = -(\eta_+ t_1 + \eta_- t_0)(1-\cos k_D ) ,\\
& =  - (\eta_+ t_1 + \eta_- t_0)\delta/ t_1  = - w_{z,\bK'_\pm } .
\end{align}
As a result, denoting the expansion of the strain perturbation as $\mathcal W_{\bK_\pm}$ and $\mathcal W_{\bK'_\pm}$ (in analogy with $\mathcal H_{\bK_\pm} $ and $\mathcal H_{\bK'_\pm} $), we have 
\be
\mathcal W_{\bK_\pm} = -w^D_0-w^D_z \tau^z, \qquad \mathcal W_{\bK'_\pm} = w^D_0+w^D_z \tau^z, \label{app:strain-Dirac-2B}
\ee
with $w^D_0$ and $w^D_z$ defined as
\begin{align}
w^D_0 &\equiv (\eta_+ t_0 + \eta_- t_1)\delta/t_1 = t_a\delta/t_1, \label{app:w_0^D} \\
w^D_z &\equiv (\eta_+ t_1 + \eta_- t_0)\delta/t_1 = t_b\delta/t_1. \label{app:w_z^D}
\end{align}
Here we have used Eqs.~\eqref{app:t_a} and \eqref{app:t_b}. The expansion of \eqref{app:strain-Dirac-2B} is the basis for determining $\Lambda$ within the continuum model. We note in passing that \eqref{app:wx}, if included in the strain coupling, does not enter the continuum model and thus does not contribute to the polarizability (at this level). 

{\it Continuum model calculation of polarizability.} A calculation of the polarizability within the continuum model is achieved in a straightforward way by applying Eq.~\eqref{eq:Lambda-2band} to each of the Dirac valleys and then summing the result. Here we proceed in two steps: we first determine the orbital piezomagnetic polarizability for the strain coupling considered in the main text, which makes the choice $t_a = t_0$ and $t_b=0$, and then determine the polarizability for the general case. As noted, in this case only the first term (i.e., ``geometric'' term) in Eq.~\eqref{eq:Lambda-2band} contributes and the polarizability directly follows from the Berry curvature and energy shifts. In this restricted case, with the Berry curvature of the Dirac points given by Eqs.~\eqref{app:Omega_K-2B} and \eqref{app:Omega_K'-2B}, and with the strain coupling given by Eq.~\eqref{app:w_0^D} (and $t_a$ set to $t_0$), one finds
\begin{align}
\Lambda  &=  \frac{e}{\hbar} \int \frac{d^2\bk}{(2\pi)^2} w^D_0 \left(\Omega^{\bK_+}_{xy}+\Omega^{\bK_-}_{xy}-\Omega^{\bK'_+}_{xy}-\Omega^{\bK'_-}_{xy}  \right) \nonumber \\
& = - \frac{e}{\pi\hbar}w^D_0 \text{sgn}(\Delta) =- \frac{e}{\pi\hbar}(t_0 \delta/t_1) \text{sgn}(\Delta) .
\end{align}
This is the result quoted and discussed in the main text.

In the case of the more general strain coupling given by Eqs.~\eqref{app:chi_k} and \eqref{app:wz_k}, and in the continuum model by Eqs.~\eqref{app:chi_k} and \eqref{app:wz_k}, the second term in Eq.~\eqref{eq:Lambda-2band} also contributes to $\Lambda$. A key observation is that for the continuum Dirac model the contribution from the second term can still be expressed in terms of the Berry curvature of the Dirac points. Specifically, for the valleys $\bK_\pm$ we find that the second term gives a contribution given by 
\be
-(\partial_y \ve^{\bK_\pm}_\bq ) \frac{\bw \cdot \bn^{\bK_\pm}_\bq \times \partial_x \bn^{\bK_\pm}_\bq}{2\left|\bn^{\bK_\pm}_\bq\right|^3} \rightarrow w^D_z \frac{v_0}{v_1} \Omega^{\bK_\pm}_{xy}.
\ee
Here we have used Eq.~\eqref{app:Dirac-Kpm} and \eqref{app:eps-n-Kpm}, and further that $\bw = -w_z^D \hat\bz$. Similarly, for the valleys $\bK'_\pm$ we find that the second term gives a contribution given by 
\be
\partial_x \ve^{\bK'_\pm}_\bq \frac{\bw \cdot \bn^{\bK'_\pm}_\bq \times \partial_y \bn^{\bK'_\pm}_\bq}{2\left|\bn^{\bK'_\pm}_\bq\right|^3} \rightarrow -w^D_z \frac{v_0}{v_1} \Omega^{\bK'_\pm}_{xy}
\ee
Collecting all contributions and using \eqref{app:top-Dirac} we then find the polarizability $\Lambda$ as
\begin{align}
\Lambda  &= - \frac{e}{\pi\hbar}\left( w^D_0 - \frac{v_0w^D_z}{v_1} \right) \text{sgn}(\Delta) \nonumber \\
& = - \frac{e}{\pi\hbar}\frac{\eta_-\delta (t^2_1 - t^2_0)}{t^2_1} \text{sgn}(\Delta).
\end{align}
In the second line we have used Eqs.~\eqref{app:w_0^D}, \eqref{app:w_z^D}, \eqref{app:t_a}, \eqref{app:t_b}, as well as the definitions of $v_0$ and $v_1$.

\section{Continuum Dirac quadrupole model of the Lieb lattice altermagnet} 
\label{app:Lieb}

This Appendix details the derivation of the continuum Dirac theory for the Lieb lattice altermagnet. The goal is in particular to demonstrate in detail how the Dirac quadrupole structure (as defined by the linearized Dirac Hamiltonians) is derived. This derivation builds on the work of Ref.~\onlinecite{Antonenko:2025p096703}, which first pointed out the existence of Dirac points in the Lieb lattice altermagnet model. 

We will proceed as follows. First, we determine the continuum Dirac model by expanding the lattice Hamiltonian in the absence of strain, and then perform a similar expansion of the strain terms, for which we only keep the lowest zeroth order contributions. Based on the linearized continuum model of the full Hamiltonian, we then show in Appendix \ref{app:Lambda-Lieb} how the piezomagnetic polarizability is computed.

\textit{Dirac quadrupole model.} It was shown in Ref.~\onlinecite{Antonenko:2025p096703} that in the limit of vanishing spin-orbit coupling linear Dirac crossings occur on the boundary of the BZ. As indicated in Fig.~\ref{fig3}(b), these Dirac points are located at $\bK_\pm = (\pi,\pi\pm k_D)$ and $\bK'_\pm = (\pi\pm k_D,\pi)$, which we refer to as the valleys. Here $k_D \in (0,\pi)$. To determine $k_D$, the equation 
\be
-2t_d(\cos k_x - \cos k_y) + \sigma N_z = 0,
\ee
is solved by substituting $(k_x,k_y) = (\pi,\pi\pm k_D)$ (for the valleys $\bK_\pm$) or $(k_x,k_y) = (\pi\pm k_D,\pi)$ (for the valleys $\bK'_\pm$). Let us consider the valleys $\bK_\pm$ located on the $k_x = \pi$ line. One then finds
\be
\sin^2\frac{k_D}{2} = \frac{\bar{\sigma}N_z}{4t_d} \quad \rightarrow \quad \sin\frac{k_D}{2} = \sqrt{\frac{\bar{\sigma}N_z}{4t_d}}, \label{app:Lieb-Dirac-condition}
\ee
from which it follows that Dirac points only occur in the $\sigma=\down$ ($\sigma=\up$) sector when $N_z>0$ ($N_z<0$). This is opposite for the $\bK'_\pm$ located on the $k_y = \pi$ line and leads to the key property of \emph{spin-valley locking}: the Dirac points at $\bK_\pm$ ($\bK'_\pm$) occur only in spin sector $\sigma=\down$ ($\sigma=\up$) \emph{or} $\sigma=\up$ ($\sigma=\down$), depending on the sign of $N_z$. We emphasize that it also follows from \eqref{app:Lieb-Dirac-condition} that the Dirac continuum model is only meaningful as long as $0< |N_z|/4t_d < 1$. The value $|N_z|=4t_d$ marks a topological transition to a trivial insulator, such that the Dirac model ceases to make sense when $|N_z|>4t_d$.

To obtain the continuum Dirac Hamiltonian, the next step is to expand the lattice Hamiltonian of Eq.~\eqref{eq:H_0-Lieb} to linear order in small momentum $\bq$ relative to the Dirac point momenta $\bK_\pm$ and $\bK'_\pm$. Consider first the piece of the dispersion given by $\ve_\bk$. Here, an expansion yields
\begin{align}
\ve_{\bK_\pm +\bq} &\approx 4t_0\cos^2\frac{k_D}{2}  \mp 2 t_0 \sin k_D q_y, \\
\ve_{\bK'_\pm +\bq} &\approx 4t_0\cos^2\frac{k_D}{2}  \mp 2 t_0 \sin k_D q_x.
\end{align}
This motivates the definition of an energy $\ve_D$ and a velocity $v_0$ as
\begin{align}
\ve_D &\equiv 4t_0\cos^2\frac{k_D}{2} = 4t_0 (1- |N_z|/4t_d), \label{app:eps_D} \\
v_0 & \equiv 2 t_0 \sin k_D = 4t_0\sqrt{\frac{|N_z|}{4t_d}}\sqrt{1-\frac{|N_z|}{4t_d}}. \label{app:v_0}
\end{align}

Next, consider the components of the vector $\bn^\sigma_\bk$, which is given by
\be
\bn^\sigma_\bk = (t_{x,\bk}, \sigma \lambda_{z,\bk}, t_{z,\bk}+\sigma N_z)
\ee
and focus first on the $x$-component, which is spin-independent and given by $t_{x,\bk}$. Expanding the $x$-component around $\bK_\pm$ and $\bK'_\pm$ yields to linear order
\begin{align}
t_{x,\bK_\pm +\bq} &\approx   \mp 2 t_1  \sin\frac{k_D}{2} q_x, \\
t_{x,\bK'_\pm +\bq} &\approx   \mp 2 t_1  \sin\frac{k_D}{2} q_y,
\end{align}
which motivates the definition of a velocity $v_1$ as
\be
v_1  \equiv 2 t_1  \sin\frac{k_D}{2} = 2t_1\sqrt{\frac{|N_z|}{4t_d}}. \label{app:v_1}
\ee
Consider then the $z$-component, which does depend on spin via the N\'eel order parameter $N_z$. Importantly, this dependence only determines in which spin sector the Dirac points at $\bK_\pm$ and $\bK'_\pm$ occur, as discussed above, and does not affect the expansion beyond the leading order constant term. Expanding the $z$-component to linear order in $\bq$, we find
\begin{align}
n^\sigma_{z,\bK_\pm +\bq} &\approx   \pm 2 t_d  \sin k_D q_y, \\
n^\sigma_{z,\bK'_\pm +\bq} &\approx   \mp 2 t_d  \sin k_D q_x,
\end{align}
which shows that there is no dependence on $\sigma$ for the $q$-linear terms. (Note the sign change between $\bK_\pm$ and $\bK'_\pm$.) This motivates the definition of a velocity
\be
v_2  \equiv 2 t_d  \sin k_D= 4t_d\sqrt{\frac{|N_z|}{4t_d}}\sqrt{1-\frac{|N_z|}{4t_d}}. \label{app:v_2}
\ee
Finally, consider the $y$-component, which is equal to $\sigma \lambda_{z,\bk}$. The lowest order term in the expansion of the function $\lambda_{z,\bk}$ is a constant, and it is therefore sufficient to expand to lowest (i.e., zeroth) order. We find
\begin{align}
\lambda_{z,\bK_\pm } &\approx   4 \lambda  \cos\frac{k_D}{2} , \\
\lambda_{z,\bK'_\pm} &\approx   4 \lambda  \cos\frac{k_D}{2} ,
\end{align}
which motivates the definition of the Dirac mass $m$ as
\be
m \equiv 4 \lambda  \cos\frac{k_D}{2} = 4 \lambda \sqrt{1-\frac{|N_z|}{4t_d}}. \label{app:m}
\ee
It is important to note that the mass $m$ enters with opposite sign at $\bK_\pm$ and $\bK'_\pm$, since $n^\sigma_{y,\bk} = \sigma \lambda_{z,\bk}$. The valleys $\bK_\pm$ and $\bK'_\pm$ correspond to opposite spin sectors and therefore have opposite sign of the mass. As noted above, to which spin sector the valleys belong depends on the sign of $N_z$, and one therefore expects a dependence on $\zeta \equiv \text{sgn}(N_z)$. 

Putting everything together yields the Dirac Hamiltonians of Eqs.~\eqref{eq:Lieb-Dirac-K} and \eqref{eq:Lieb-Dirac-K'}. As mentioned in the main text, these Dirac Hamiltonians describe tilted and gapped Dirac cones, with a tilt that originates from the expansion of $\ve_\bk$. The tilt velocity is given by $v_0$ defined in Eq.~\eqref{app:v_0}. 

\textit{Berry curvature.} It is straightforward to determine the Berry curvature $\Omega_{xy}$ of the occupied valence bands in each valley. We find
\begin{align}
\Omega^{\bK_\pm}_{xy}& = -\zeta\frac{m v_1 v_2}{2(m^2 +v_1^2q_x^2+v_2^2q_y^2)^{3/2}}, \\
\Omega^{\bK'_\pm}_{xy} &= \zeta\frac{m v_1 v_2}{2(m^2 +v_1^2q_y^2+v_2^2q_x^2)^{3/2}} ,
\end{align}
which confirms the Dirac quadrupole structure of the four Dirac valleys. 

\textit{Strain coupling.} Next, we determine the continuum model expansion of the strain coupling. In the main text we introduced the general form of the strain coupling as
\be
W_\bk = \chi_\bk + w_{z,\bk}\tau^z, \label{app:W_k-Lieb}
\ee
with the functions $\chi_\bk$ and $ w_{z,\bk}$ given by
\begin{align}
\chi_\bk & = -2t_0 (\cos k_x - \cos k_y) ,\label{app:chi-Lieb} \\
w_{z,\bk} & = -2t_d (\cos k_x + \cos k_y). \label{app:wz-Lieb}
\end{align}
These functions capture anisotropic next-nearest neighbor (and intra-sublattice) hoppings allowed in the presence of strain. In the continuum model expansion we only require the lowest order terms, which are constant $\bq$-independent terms obtained from evaluating $\chi_\bk$ and $ w_{z,\bk}$ at $\bK_\pm$ and $\bK'_\pm$. We find
\be
\chi_{\bK_\pm } =- \chi_{\bK'_\pm }= 2t_0 (1 + \cos k_D) = 4t_0 \cos^2\frac{k_D}{2}, \label{app:chi-valley-Lieb}
\ee
and 
\be
w_{z,\bK_\pm } =w_{z,\bK'_\pm} = 2t_d (1 - \cos k_D)=4t_d \sin^2\frac{k_D}{2}, \label{app:wz-valley-Lieb}
\ee
such that the terms in the continuum Dirac model coming from the strain coupling, which we denote $\mathcal W_{\bK_\pm}$ and $\mathcal W_{\bK'_\pm}$, can be expressed as
\be
\mathcal W_{\bK_\pm} = w^D_0 + w^D_z \tau^z, \quad \mathcal W_{\bK'_\pm} = -w^D_0 + w^D_z \tau^z, \label{app:Dirac-strain-Lieb}
\ee
with the constants $w^D_0$ and $w^D_z$ defined as
\be
w^D_0 \equiv 4 t_0 \left(1-\frac{|N_z|}{4t_d}\right), \quad w^D_z \equiv 4 t_d \frac{|N_z|}{4t_d}. \label{app:w0-wz}
\ee
Note that the full linearized continuum model Hamiltonian in each valley then takes the form $\mathcal H_{\bK_\pm}+\phi \mathcal W_{\bK_\pm}$ and $\mathcal H_{\bK'_\pm}+\phi \mathcal W_{\bK'_\pm}$.

\mbox{}

\section{Calculation of piezomagnetic polarizability for the Lieb lattice altermagnet} 
\label{app:piezomagnet-Lieb}

This Appendix provides more details on the calculation of the orbital piezomagnetic polarizability in the Lieb lattice model. 

\textit{Full expression of piezomagnetic polarizability.} As stated in the main text, the piezomagnetic polarizability of the Lieb lattice model is a sum over the contributions from each spin sector, with each of these contributions being described by \eqref{eq:Lambda-2band}. The explicit expression for $\Lambda$ is then
\begin{multline}
\Lambda = 
- \frac{e}{\hbar} \sum_\sigma \int \frac{d^2\bk}{(2\pi)^2}\bigg\{ \chi_\bk \frac{\bn^\sigma_\bk \cdot \partial_x \bn^\sigma_\bk \times \partial_y \bn^\sigma_\bk}{2|\bn^\sigma_\bk|^3}  \\
 +\bigg[ \partial_x \varepsilon_\bk \frac{\bw_\bk \cdot \bn^\sigma_\bk \times \partial_y \bn^\sigma_\bk}{2|\bn^\sigma_\bk|^3} -  (x\leftrightarrow y) \bigg] \bigg\}, \label{app:Lambda-Lieb}
\end{multline}
with $\bn^\sigma_\bk$ and $\varepsilon_\bk$ defined in Eq.~\eqref{eq:H_0-Lieb}.

\textit{Piezomagnetic polarizability within continuum model.} Within the linearized Dirac continuum model, the polarizability can be calculated by applying the formula \eqref{app:Lambda-Lieb} to each of the four valleys, with the understanding that the valleys are spin-valley locked. Here we choose the valley $\bK_+$ as an example. 

For the valley $\bK_+$, we can replace $\chi_\bk$ by $w^D_0$ (see Appendix~\ref{app:Lieb}) and we can take the vector $\bn = \bn_\bq$ as $\bn = (-v_1q_x,-\zeta m , v_2 q_y)$. The vector $\bw$ is simply given by $\bw = (0,0,w^D_z)$. With these ingredients we evaluate both the geometric and the interband terms in \eqref{app:Lambda-Lieb} (i.e., the first and second terms in the integrand). From simple inspection we see that the geometric term is proportional to the Berry curvature of the occupied valence band, which is multiplied by the energy shift $w^D_0$. In the special case of the linearized Dirac model, the interband term can also be rewritten in terms of the Berry curvature (by multiplying and dividing by $v_2$), while $\partial_y \varepsilon_{\bK_+ + \bq} $ becomes $\partial_y (-v_0q_y) = -v_0$. Putting everything together, one obtains
\begin{align}
\Lambda_{\bK_+}&= -\frac{e}{2\pi \hbar} \int\frac{d^2\bq}{2\pi}\left(w^D_0 + \frac{v_0}{v_2}w^D_z \right) \Omega^{\bK_+}_{xy} \\
& = \zeta \text{sgn}(m)\frac12 \frac{e}{2\pi \hbar} \left(w^D_0 + \frac{v_0}{v_2}w^D_z \right)
\end{align}
for the contribution from the valley $\bK_+$. Since the sign of the mass $m$ is determined by the spin-orbit coupling $\lambda$, we can replace $\text{sgn}(m)$ by $\text{sgn}(\lambda)$. Furthermore, we can use $\zeta = \text{sgn}(N_z) $ and write $\zeta \text{sgn}(\lambda) = \text{sgn}(\lambda N_z)$ 

It is now a straightforward matter to determine the full polarizability $\Lambda$ by considering the contributions from the other valleys. The valley $\bK_-$ yields the same geometric contribution, since both the Berry curvature and the energy shift $w_0$ are the same. Furthermore, since the sign of both velocities $v_0$ and $v_1$ changes, the interband contribution is also the same. 

The two valleys $\bK'_\pm$ also yield the same geometric contribution, since both the Berry curvature and the energy shift $w_0$ change sign. The interband contribution is also the same, since the sign change of the Berry curvature is compensated by the relative sign change between the $\partial_x \varepsilon_\bk $ and $\partial_y \varepsilon_\bk $ terms. As a result, we find that the piezomagnetic polarizability computed within the continuum model is given by
\be 
\Lambda=  2 \text{sgn}(\lambda N_z) \frac{e}{2\pi \hbar} \left(w^D_0 + \frac{v_0}{v_2}w^D_z \right).
\ee
This expression can be simplified by using the definitions of $w^D_0$, $w^D_z$, $v_0$, and $v_2$, as given by Eqs.~\eqref{app:w0-wz}, \eqref{app:v_0}, and \eqref{app:v_2}. In particular, using that $v_0/v_2 = t_0/t_d$ we find the simple result
\be
\left(w^D_0 + \frac{v_0}{v_2}w^D_z \right) = 4t_0.
\ee
Dividing and multiplying by $t_1$, and replacing $t_1 = \hbar^2/2m_e a^2$, we obtain
\be
\Lambda= \text{sgn}(\lambda N_z) \frac{4 t_0}{\pi t_1}\times \frac{\mu_B}{a^2}.
\ee

\section{Topological transition in the Lieb lattice altermagnet} 
\label{app:Lieb-transition}

This Appendix provides an analysis of the Lieb lattice piezomagnetic polarizability as a function of the N\'eel order parameter $N_z$, focusing in particular on the topological transition at $|N_z|=N_{z,c} \equiv 4t_d$. Here $N_{z,c}$ is defined as the critical value of $N_z$ at which the transition occurs.

\begin{figure}
	\includegraphics[width=\columnwidth]{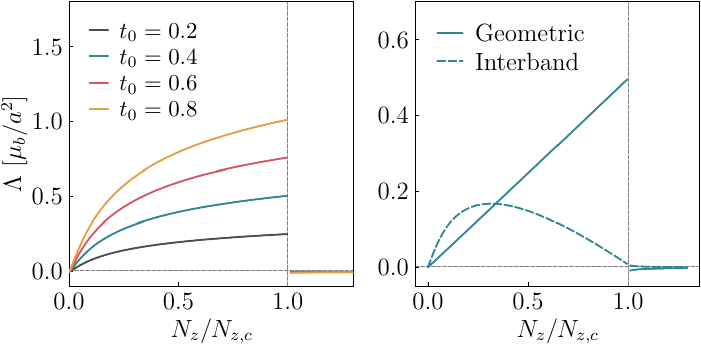}
	\caption{(a) Piezomagnetic polarizability $\Lambda$ of the Lieb lattice model as a function of $N_z$ (in units of $N_{z,c}\equiv 4t_d$) for different $t_0$ (in units of $t_1$). We have set $(t_d,\lambda) = (3t_1,0.2t_1)$. (b) Same as in (a), showing separate contributions from the geometric and interband terms in Eq.~\eqref{eq:Lambda-2band} (for $t_0=0.4t_1$). }
	\label{fig4}
\end{figure}

In Fig.~\ref{fig4}(a) we show the polarizability $\lambda$ as a function of the magnetic order parameter $N_z$, in units of $N_{z,c}$. A key feature is the discontinuity of the polarizability right at the topological transition, when $N_z = N_{z,c}$ (here we choose $N_z > 0$ for concreteness). At this topological transition the gap closes and the system becomes a trivial insulator with vanishing mirror Chern number. The gap closing transition is described by a Dirac fermion mass inversion at both $X=(\pi,0)$ and $Y=(0,\pi)$. The discontinuity of $\Lambda$ can be captured by a linearized continuum model obtained from expanding the lattice Hamiltonian in small momentum $\bq$ around the $X$ and $Y$ points. Since here we focus on $N_z > 0$ for concreteness, it is straightforward to determine that the gap closing transition at $X$ ($Y$) is in the $\sigma=\down$ ($\sigma=\up$) sector.

The expansion of the Lieb lattice Hamiltonian around the $X$ and $Y$ points follows the same recipe as the expansion described in Appendix~\ref{app:Lieb}. Expanding the function $t_{x,\bk}$ yields
\begin{align}
t_{x,X +\bq} &\approx    2 t_1  q_x, \\
t_{x,Y +\bq} &\approx    2 t_1  q_y,
\end{align}
while expanding the function $\lambda_{z,\bk}$ gives
\begin{align}
\lambda_{z,X +\bq} &\approx  2 \lambda  q_y , \\
\lambda_{z,Y +\bq} &\approx   2 \lambda q_x .
\end{align}
This motivates the definition of two velocities $v'_1$ and $v'_2$ as
\be
v'_1 \equiv 2t_1, \qquad v'_2 \equiv 2\lambda.
\ee
To proceed, we write $N_z$ as $N_z = N_{z,c}+ m $, such that $m$ can be viewed as a band inversion parameter; $m  < 0$ describes the topological (mirror Chern insulator) phase, whereas $m  > 0$ describes the trivial phase. With $N_z$ written in this way, the linearized continuum model Hamiltonian at $X$ and $Y$ takes the form
\begin{align}
\mathcal H_{X } &=  v'_1 q_x \tau^x- v'_2 q_y \tau^y - m \tau^z, \label{app:H_X} \\
\mathcal H_{Y} &=  v'_1 q_y \tau^x+ v'_2 q_x \tau^y + m\tau^z ,\label{app:H_Y}
\end{align}
showing that $m$ enters as a mass term, as expected for a theory which describes a topological transition. Note that there are no linear-in-q terms proportional to the identity at $X$ and $Y$, such that the Dirac cones corresponding to $\mathcal H_{X }$ and $\mathcal H_{Y }$ are not tilted. As a result, when computing the polarizability $\Lambda$ from the continuum model only the geometric term will contribute; the contribution from the interband term vanishes.

To compute the polarizability within the continuum model, we need to evaluate $\chi_{\bk}$ at the $X$ and $Y$ points. This determines how the Dirac points are shifted in energy in response to strain. We find
\be
\chi_{X } \approx  4t_0 , \qquad
\chi_{Y } \approx  -4t_0 ,
\ee
and from this it is straightforward to obtain
\begin{align}
\Lambda  &=  -\frac{e}{\hbar} \int \frac{d^2\bk}{(2\pi)^2} 4t_0 \left(\Omega^{X}_{xy}-\Omega^{Y}_{xy} \right) \nonumber \\
& = - \frac{2e}{\pi\hbar}t_0\text{sgn}(m).
\end{align}
Here we have introduced $\Omega^{X}_{xy}$ and $\Omega^{Y}_{xy}$ as the Berry curvatures at $X$ and $Y$, which are straightforwardly obtained from Eqs.~\eqref{app:H_X} and \eqref{app:H_Y}. 

From this result it follows that at $m=0$ the polarizability jumps by an amount $4t_0/\pi t_1$, in units of $\mu_B /a^2$. This is precisely what we find in Fig.~\ref{fig4}: the discontinuity at $m=0 $ found in the full lattice calculation is in perfect agreement with the result from the Dirac model calculation. Furthermore, as mentioned, in the continuum model calculation only the geometric term contributes to the polarizability, which is confirmed by the numerical lattice model calculation shown in Fig.~\ref{fig4}(b). In Fig.~\ref{fig4}(b) we separately show the contributions from the geometric and the interband terms, revealing that only the geometric term exhibits a discontinuity at the topological transition.

\end{document}